\documentclass[10pt]{article}

\usepackage{epsf}
\usepackage{amsmath}
\usepackage{amssymb}
\usepackage{graphicx}
\usepackage{dcolumn}
\usepackage{bm}
\usepackage{color}

\begin{document}

\def\thefootnote{\fnsymbol{footnote}}

\begin{center}
\Large\bf\boldmath
Cosmological constraints on unifying Dark Fluid models\unboldmath
\end{center}
\vspace{0.6cm}
\begin{center}
A. Arbey\footnote{Electronic address: \tt arbey@obs.univ-lyon1.fr}\\[0.4cm]
{\sl Universit\'e de Lyon, Lyon, F-69000, France ; Universit\'e Lyon~1,
Villeurbanne, F-69622, France ; Centre de Recherche Astrophysique de
Lyon, Observatoire de Lyon, 9 avenue Charles Andr\'e, Saint-Genis Laval cedex, F-69561, France ; CNRS, UMR
5574 ; Ecole Normale Sup\'erieure de Lyon, Lyon, France}
\end{center}
\vspace{0.6cm}
\begin{abstract}
\noindent In the standard model of cosmology, dark matter and dark energy are presently the two main
contributors to the total energy in the Universe. However, these two dark components are still of unknown
nature, and many alternative explanations are possible. We consider here the so-called unifying dark fluid
models, which replace dark energy and dark matter by a unique dark fluid with specific properties. We will
analyze in this context recent observational data from supernov\ae~of type Ia, large scale structures and
cosmic microwave background, as well as theoretical results of big-bang nucleosynthesis, in order to derive
constraints on the dark fluid parameters. We will also consider constraints from local scales, and conclude
with a brief study of a scalar field dark fluid model.
\end{abstract}
\vspace{0.3cm}
\section{Introduction}
\noindent In the standard model of cosmology, the total energy density of the Universe is dominated today by the densities of two components: the first one, called ``dark matter'', is generally modeled as a system of collisionless particles (``cold dark matter'') and consequently has an attractive gravitational effect like usual matter. The second one, generally refereed as ``dark energy'' or ``cosmological constant'' can be considered as a vacuum energy with a negative pressure, which seems constant today. The real nature of these two components remains unknown, and the proposed solutions to these problems are presently faced to numbers of difficulties.\\
The general approach to try to identify these two distinct components is to find constraints to better understand their behaviors. The usual method to find the best values of the different parameters of the model is to predict observations and to adjust the parameters to improve the accuracy of the predictions. In this paper, I consider the so-called dark fluid model \cite{arbey3}, in which the dark matter and the dark energy are in fact different aspects of a same dark fluid. For simplicity reasons, it is considered in the following that this fluid is perfect, {\it i.e.} the entropy variations and the shear stress can be ignored. I will examine data from the latest observations of supernov\ae~of type Ia, cosmic microwave background (CMB), large scale structures, and the theoretical predictions from big-bang nucleosynthesis (BBN), and I will derive constraints on the dark fluid parameters. I will then describe the necessary behavior of a dark fluid at local scales. In a last paragraph, I will discuss the scalar field dark fluid model first presented in \cite{arbey3}.
\section{Constraints from the Supernovae of Type Ia}
\noindent Cosmological constraints from the supernov\ae~of type Ia are based on the joint observations of the redshift $z$ and of the apparent luminosity $l$ of a large number of supernov\ae~\cite{SNeIa1}. If only one component exists -- replacing the two dark components -- it should lead to the same influence on the luminosity distance -- and then on the expansion of the Universe -- than dark matter and dark energy would have. It seems clear that if:
\begin{eqnarray}
\rho_D=\rho_{dm}+\rho_{de} \;\; , \;\;P_D=P_{dm}+P_{de}=P_{de} \;\; ,
\end{eqnarray}
where $dm$, $de$ and $D$ stands respectively for dark matter, dark energy and dark fluid, and where $P$ and $\rho$ are the pressures and densities of the components, the dark fluid would provide the same effect on the expansion of the Universe as the two components.\\
Observations on the supernov\ae~of type Ia provide constraints on the dark component densities and on the dark energy behavior \cite{SNeIa2}. From these constraints, it is possible to characterize the dark fluid at low redshifts. First, the cosmological parameter corresponding to the dark fluid can be written in function of those related to dark matter and dark energy:
\begin{equation}
\Omega_D^0 = \Omega_{dm}^0 + \Omega_{de}^0 \;\; .
\end{equation}
The equations of state of dark energy and dark fluid can be written \cite{eos}
\begin{equation}
\omega \equiv \displaystyle \frac{P}{\rho} = \omega^0 + (1-a) \omega^a
\end{equation}
where $a$ is the scale factor and $\omega_0$ and $\omega_a$ can be considered as two constants. At first order in the redshift $z$, they become for low redshifts:
\begin{equation}
\omega = \omega^0 + \omega^a z \;\;.
\end{equation}
We have observational constraints on the values of $\omega_{de}^0$ and $\omega_{de}^a$, and would like to deduce constraints on $\omega_D^0$ and $\omega_D^a$. The equation of state of the dark fluid writes:
\begin{equation}
\omega_D=\frac{P_D}{\rho_D}=\frac{P_{de}}{\rho_{dm}+\rho_{de}}=\omega_{de} \frac{\rho_{de}}{\rho_{dm}+\rho_{de}} \;\;.
\end{equation}
The dark matter density evolves like $\rho_{dm}=\rho^0_{dm} a^{-3} = (1+3z) \rho^0_{dm}$ . To study the behavior of $\rho_{de}$, let us assume that at low redshift:
\begin{equation}
\rho_{de}=\rho_{de}^0+z \rho_{de}^1 \;\;.
\end{equation}
The equation of conservation of the energy-momentum tensor for dark energy satisfies:
\begin{equation}
\frac{d}{dt}(\rho_{de} a^3) = -P_{de} \frac{d(a^3)}{dt} \;\;,
\end{equation}	
which becomes at low redshift:
\begin{equation}
\frac{d}{dt}(\rho_{de}^0 + z (\rho_{de}^1 - 3\rho_{de}^0)) = - (\omega_{de}^0
\rho_{de}^0) \frac{d(1-3z)}{dt} \;\;,
\end{equation}
so that the density of dark energy reads:
\begin{equation}
\rho_{de}^1 = 3 \rho_{de}^0 (1+\omega_{de}^0) \;\;.
\end{equation}
Then, the relation between the ratio pressure/density for the dark fluid becomes:
\begin{equation}
\omega_D=(\omega_{de}^0+\omega_{de}^a z) \frac{\rho_{de}^0 (1 + 3 (1+\omega_{de}^0) z)}{(1+3z) \rho_{dm}^0+\rho_{de}^0 (1 + 3 (1+\omega_{de}^0) z)}\;\;,
\end{equation}
and one can determine the value of the two first terms of the expansion:
\begin{eqnarray}
\omega_D^0= \frac{\omega_{de}^0 \Omega^0_{de}}{\Omega^0_{dm} + \Omega^0_{de}}\;\;,\;\; \omega_D^a= \frac{\omega_{de}^a \Omega^0_{de}}{\Omega^0_{dm} + \Omega^0_{de}} + \frac{3 \Omega^0_{dm}\Omega^0_{de} (\omega_{de}^0)^2}{(\Omega^0_{dm} + \Omega^0_{de})^2}\;\;.
\end{eqnarray}
The favored values for the cosmological parameters of the usual standard model from the supernov\ae~of type Ia \cite{SNeIa1} combined with the results of other cosmological observations \cite{wmap5} lead to the following parameters for the dark fluid:
\begin{eqnarray}
\Omega_D^0 = 1.005 \pm 0.006  \;\;,\;\;\omega_D^0 = -0.80 \pm 0.12 \;\;,\;\; \omega_D^a =  0.9 \pm 0.5 \;\;.
\label{resultSN}
\end{eqnarray}
Recent supernova observations confirm that dark energy has negative pressure. Moreover, $\omega_{de}<-1$ is not excluded, and in this case the dark energy cannot be explained anymore thanks to the usual scalar field models (see for example \cite{caldwell} for a possible way out of this problem). One can see that this difficulty vanishes with a dark fluid. Hence, the pressure of the fluid has to be negative today at cosmological scales, and seems to increase strongly with the redshift. As $\omega_D^0 \ge -1$, it is possible to model for example the dark fluid with a scalar field.
\section{Large Scale Structures}
\noindent To study the large scale structure formation conditions, we consider that structure formation in the usual standard model is dominated by dark matter, and that dark energy has only a late influence. We can therefore assume here that the equation of state of the fluid does not change during the growth of perturbations. To simplify, we also consider that the entropy perturbations can be ignored and that the Jeans length is smaller than any other considered scale.\\
Following \cite{LSS}, in the fluid approximation, one can write the evolution equation of the local density contrast:
\begin{equation}
\label{eq.evol}
\displaystyle \frac{1}{H^2}\frac{d^2\delta}{dt^2} + \left(2+ \frac{\dot{H}}{H^2}\right) \frac{1}{H}\frac{d\delta}{dt} - \frac{2}{3} (1+\omega_D)(1+3\omega_D) \Omega_D \delta  = \frac{4+3\omega_D}{3(1+\omega_D)} \frac{1}{1+\delta} \left( \frac{1}{1+\delta}\right)^2 \frac{1}{H^2} \left( \frac{d\delta}{dt} \right)^2 + \frac{3}{2} (1+\omega_D) (1+3\omega_D) \Omega_D \delta^2
\end{equation}
where $\displaystyle \delta\left(\vec{x},t\right) \equiv \frac{\rho_D\left(\vec{x},t)\right)}{\overline{\rho_D}(t)} - 1 $ is the density contrast. If the dark fluid is completely dominant at the time of the growth of structures, one can show that equation (\ref{eq.evol}) admits the solution, in the linear approximation:
\begin{equation}
\delta\left(\vec{x},t\right)= \delta_1\left(\vec{x}\right) a^{1 + 3 \omega_D} + \delta_2\left(\vec{x}\right) a^{- \frac{3}{2} (1+\omega_D)} \;\;.
\end{equation}
Therefore one has only a growing mode if $\omega_D > -1/3$ or if $\omega_D < -1$.\\
\\
To keep our analysis general, we will not consider further the large scale structure formation conditions, as it would require the specification of a precise dark fluid model and the study of a complete scenario of structure formation.
\section{Cosmic Microwave Background}
\noindent The new five-year WMAP results have been published recently. The precision of the temperature power spectrum which has been deduced from the observations of the cosmic microwave background (CMB) has been greatly improved compared to the previous results. Such a power spectrum can be predicted within the cosmological standard model thanks to a program such as CMBFAST \cite{cmbfast}. In this section, we will estimate the observed positions of the peaks and compare them to the observations to constrain the parameters of the dark fluid.\\
First, one can note that at high redshift, in the standard cosmological model the density of dark energy is nearly negligible as comparison to the density of the dark matter. As the standard cosmological model is able to correctly reproduce the fluctuations of the CMB, we can assume that our dark fluid should not behave very differently from the superposition of dark matter and dark energy, and so should behave at the moment of recombination nearly like matter. Therefore, we can write the density of the fluid as a sum of a matter--like term ($m$) and of another term of unknown behavior ($u$):
\begin{equation}
\rho_D = \rho_{Dm}^{ls} \left(\frac{a}{a_{ls}}\right)^{-3} +  \rho_{Du} \;\;.
\end{equation}
where $ls$ stands for \textit{last scattering}. We do not want to specify a model of dark fluid in order to be as general as possible. Nevertheless, we will consider for simplicity only the background properties of the dark fluid. The conformal time is defined by:
\begin{equation}
\tau = \int dt \,\, a^{-1}(t) \;\;.
\end{equation}
The spacing between the peaks is then given, to a good approximation, by \cite{hu_sugiyama}:
\begin{equation}
\Delta l \approx \pi \frac{\tau_0 - \tau_{ls}}{\overline{c}_s \tau_{ls}} \;\;, \label{deltaldef}
\end{equation}
where $\overline{c}_s$ is the average sound speed before last scattering, and $\tau_0$ and $\tau_{ls}$ the conformal times today and at last scattering:
\begin{equation}
\overline{c}_s \equiv \tau_{ls}^{-1} \int_0^{\tau_{ls}} d\tau \left( 3 + \frac{9 \rho_b(t)}{4 \rho_r(t)}\right)^{-1/2} \;\;,
\end{equation}
where $\rho_b$ is the density of baryonic matter and $\rho_r$ is the density of relativistic fluids (radiation and neutrinos).\\
\\
For a flat Universe, the Friedmann equations becomes:
\begin{equation}
H^2= H_0^2 \left( \Omega^0_b a^{-3} + \Omega^0_r a^{-4} + \Omega^{ls}_{Dm} \left(\frac{a}{a_{ls}}\right)^{-3} \right) + \frac{8\pi G}{3} \rho_{Du} \;\;.
\end{equation}
This equation cannot be solved without specifying the form of $\rho_{Du}$. In our case, one can assume that the fraction
\begin{equation}
\Omega_{Du} (\tau) \equiv \frac{\rho_{Du}(\tau)}{\sum \rho(\tau)}
\end{equation}
does not vary too rapidly before the moment of last scattering, so that an effective average can be defined:
\begin{equation}
\overline{\Omega}^{ls}_{Du} \equiv \tau_{ls}^{-1} \int_0^{\tau_{ls}} \Omega_{Du} (\tau) d\tau \;\;.
\end{equation}
Therefore in the following, we will consider the effective density:
\begin{equation}
\rho_{Du} \approx H^2 \frac{3}{8\pi G} \overline{\Omega}^{ls}_{Du} \;\;.
\end{equation}
After inserting this density into the Friedmann equation and replacing usual time by conformal time, the Friedmann equation becomes:
\begin{equation}
\left( \frac{da}{d\tau} \right)^2  = H_0^2 (1 - \overline{\Omega}^{ls}_{Du})^{-1} \left( ( \Omega^0_b + \Omega^{ls}_{Dm} a_{ls}^3) a(\tau)+ \Omega^0_r \right) \;\;.
\end{equation}
Solving this equation gives the value of the conformal time at the moment of last scattering:
\begin{equation}
\tau_{ls} = 2 H_0^{-1} \sqrt{\frac{1 - \overline{\Omega}^{ls}_{Du}}{\Omega^0_b + \Omega^{ls}_{Dm} a_{ls}^3}} \left\{ \sqrt{a_{ls} + \frac{\Omega^0_r}{\Omega^0_b +
\Omega^{ls}_{Dm} a_{ls}^3}} - \sqrt{\frac{\Omega^0_r}{\Omega^0_b +
\Omega^{ls}_{Dm} a_{ls}^3}}\right\} \;\;. \label{tauls}
\end{equation}
In what follows the same method is used to evaluate the conformal time today. The Friedmann equation reads, after last scattering:
\begin{equation}
\left( \frac{da}{d\tau} \right)^2 = H_0^2 \left( \Omega^0_b a(\tau) + \Omega^0_r + a(\tau)^4 \;\; \frac{\rho_{D}}{\rho_0^C}\right)\;\;,
\end{equation}
where $\rho_0^C$ is the critical density. Let us make now consider that the dark fluid has the effective equation of state:
\begin{equation}
\rho_D = \tilde{\rho}_D^0 a^{-3(1+\overline{\omega}_D)} \;\;,
\end{equation}
where $\tilde{\rho}_D^0$ is an effective value of the dark fluid density such as
\begin{equation}
\tilde{\rho}_D^0 = a_{ls}^{3(1+\overline{\omega}_D)} \rho_D^{ls} \;\;,
\end{equation}
and where $\overline{\omega}_D$ is the average value of $\omega_D$ over the conformal time
\begin{equation}
\overline{\omega}_D \equiv \frac{\displaystyle \int_0^{\tau_0} \Omega_D(\tau) \omega_D(\tau) d\tau}{\displaystyle \int_0^{\tau_0} \Omega_D(\tau) d\tau} \;\;, \label{omegaDmoy}
\end{equation}
with
\begin{equation}
\Omega_D(\tau) = \frac{\rho_D(\tau)}{\sum \rho(\tau)} \;\;.
\end{equation}
The presence of the weight $\Omega_D(\tau)$ in Eq. (\ref{omegaDmoy}) reflects the fact that the equation of state of the fluid should be more significant when its density contributes more to the total density of the Universe.\\
Defining the effective cosmological parameter $\tilde{\Omega}^0_D = 8\pi G \tilde{\rho}^0_D / (3 H_0^2)$, the Friedmann equation becomes:
\begin{equation}
\label{eq.omegatilde}
\left( \frac{da}{d\tau} \right)^2 = H_0^2 \left( \Omega^0_b a(\tau) + \Omega^0_r + \tilde{\Omega}^0_D a(\tau)^{(1-3\overline{\omega}_D)} \right) \;\;.
\end{equation}
One can then integrate this equation, and show that:
\begin{equation}
\tau_{0} = 2 H_0^{-1} F(\overline{\omega}_D) \;\;, \label{tau0}
\end{equation}
with
\begin{equation}
F(\overline{\omega}_D) \equiv \frac{1}{2} \int_{a_{ls}}^1 da \left( \Omega^0_b a + \Omega^0_r + \tilde{\Omega}^0_D a^{(1-3\overline{\omega}_D)} \right)^{-1/2} \;\;.
\end{equation}
There is no analytical solution for this function, but in a few cases. Substituting Eqs. (\ref{tauls}) and (\ref{tau0}) into Eq. (\ref{deltaldef}) we get the spacing between the peaks:
\begin{equation}
\Delta l=\pi \overline{c}_s^{-1} \left[ F(\overline{\omega}_D) \sqrt{\frac{\Omega^0_b + \Omega^{ls}_{Dm} a_{ls}^3}{1 - \overline{\Omega}^{ls}_{Du}}} \left\{ \sqrt{a_{ls} + \frac{\Omega^0_r}{\Omega^0_b +
\Omega^{ls}_{Dm} a_{ls}^3}} - \sqrt{\frac{\Omega^0_r}{\Omega^0_b +
\Omega^{ls}_{Dm} a_{ls}^3}}\right\}^{-1} -1\right] \;\;. \label{deltal}
\end{equation}
The sound velocity $\overline{c}_s$ is given by:
\begin{equation}
\overline{c}_s = \tau_{ls}^{-1} H_0^{-1} \sqrt{1 - \overline{\Omega}^{ls}_{Du}} \int_0^{a_{ls}} da \left[ \left(3 + \frac{9 \Omega_b^0}{4 \Omega_r^0} a \right)  \left( ( \Omega^0_b + \Omega^{ls}_{Dm} a_{ls}^3) a+ \Omega^0_r \right)\right]^{-1/2} \;\;.
\end{equation}
This equation can be integrated analytically, and one finally obtains:
\begin{eqnarray}
\overline{c}_s &&= \frac{1}{3} \left(\frac{\Omega_r^0}{\Omega^0_b}\right)^{1/2} \left\{ \sqrt{a_{ls} + \frac{\Omega^0_r}{\Omega^0_b +
\Omega^{ls}_{Dm} a_{ls}^3}} - \sqrt{\frac{\Omega^0_r}{\Omega^0_b +\Omega^{ls}_{Dm} a_{ls}^3}} \right\}^{-1} \times\\
\nonumber&&\ln\Biggl(\frac{\Omega_r^0 (7\Omega_b^0+4\Omega^{ls}_{Dm} a_{ls}^3) +6 a_{ls}\Omega^0_b(\Omega^0_b + \Omega^{ls}_{Dm} a_{ls}^3)+ 2 \sqrt{3\Omega_b^0 (\Omega^0_b + \Omega^{ls}_{Dm} a_{ls}^3)(\Omega^0_b a_{ls}+
\Omega^{ls}_{Dm} a_{ls}^4 + \Omega^0_r)(3 \Omega^0_b a_{ls} + 4\Omega^0_r)}}{\Omega_r^0 (7\omega_b^0+4\Omega^{ls}_{Dm} a_{ls}^3)+ 4 \Omega_r^0 \sqrt{3\Omega_b^0 (\Omega^0_b +\Omega^{ls}_{Dm} a_{ls}^3)}} \Biggr) \;\;.
\end{eqnarray}
To determine the value of $a_{ls}$, an approximate formula can be taken from~\cite{hu_fukugita}:
\begin{equation}
a_{ls}^{-1} \approx 1008 (1+0.00124 (\Omega_b^0 h^2)^{-0.74}) (1+c_1 (\Omega^{ls}_{Dm} a_{ls}^3)^{c_2}) \;\;,
\end{equation}
where
\begin{eqnarray}
c_1 = 0.0783 (\Omega_b^0 h^2)^{-0.24}(1+39.5 (\Omega_b^0 h^2)^{0.76})^{-1}\;\;,\;\; c_2 = 0.56 (1+21.1 (\Omega_b^0 h^2)^{1.28})^{-1} \;\;.
\end{eqnarray}
At this step, we know all the necessary parameters. There is a direct dependence between $\Delta l$ and the parameters $\Omega^{ls}_{Dm}$, $\overline{\Omega}^{ls}_{Du}$, $\tilde{\Omega}_D^0$ and $\overline{\omega}_D$. Even if Eq. (\ref{deltal}) is only an approximate formula, it gives the possibility to study directly the effects of the dark fluid on the position of the peaks, provided its behavior does not differ much from the requirements of the approximations on $\overline{\Omega}^{ls}_{Du}$ and $\overline{\omega}_D$.\\
However, the calculated $\Delta l$ cannot be directly related to the observed spacing between peaks, as shifts of peaks can be induced by other effects as well. In particular, the location of the i--th peaks can be approximated by:
\begin{equation}
l_i=\Delta l (m-\phi_i)=\Delta l (m-\bar\phi-\delta\phi_i) \;\;,
\end{equation}
where $\bar\phi$ is the shift of the first peak, corresponding to an overall shift, and the $\delta\phi_i$ is the specific shift of the i--th peak. Transcribing the fitting formulae of \cite{doran_lilley} in the dark fluid model, one can determine analytically the positions of the peaks, and compare our results to the data. The Five-Year WMAP data \cite{wmap5}, together with the data by BOOMERang \cite{boomerang} and ACBAR \cite{acbar}, provide the precise location of the first peaks:
\begin{eqnarray}
l_{p_1} = 219.8 \pm 0.5\;\;, \;\; l_{p_2} = 537 \pm 4 \;\;,\;\; l_{p_3} = 818 \pm 3 \;\;.
\end{eqnarray}
To evaluate roughly the value of the parameters of the fluid, one can fix the other parameters as follows:
\begin{eqnarray}
n = 1\;\;, \;\;h = 0.70\;\;, \;\;\Omega_b^0 = 0.046\;\;, \;\;\Omega_r^0 = 9.89 \times 10^{-5} \;\;.
\end{eqnarray}
\noindent Matching the observed positions of the peaks and the calculated positions reveals that:
\begin{eqnarray}
&\Omega^{ls}_{Dm} \in [0.30,0.50] & \overline{\Omega}^{ls}_{Du} \le \in [0.03,0.10] \label{resultCMB}\\
&\tilde{\Omega}_D^0 \in [0.15,1.10] & \overline{\omega}_D \in [-0.15,0.15]\nonumber
\end{eqnarray}%
\begin{table*}
\begin{center}
\begin{tabular}{cccc} \hline\hline
$\tilde{\Omega}_D^0$ & $\Omega^{ls}_{Dm}$ &  $\overline{\omega}_D$ & $\overline{\Omega}^{ls}_{Du}$\\
\hline
~~~1.10~~~ & ~~~0.45~~~ & ~~~-0.16~~~ & ~~~0.10~~~\\
1.00 & 0.40 & -0.16 & 0.08\\
0.95 & 0.40 & -0.15 & 0.09\\
0.85 & 0.35 & -0.15 & 0.05\\
0.75 & 0.40 & -0.11 & 0.09\\
0.70 & 0.40 & -0.10 & 0.07\\
0.65 & 0.30 & -0.12 & 0.04\\
0.60 & 0.35 & -0.09 & 0.05\\
0.50 & 0.40 & -0.04 & 0.08\\
0.45 & 0.35 & -0.04 & 0.04\\
0.40 & 0.40 & 0.00 & 0.08\\
0.35 & 0.45 & 0.04 & 0.09\\
0.30 & 0.45 & 0.07 & 0.09\\
0.25 & 0.40 & 0.09 & 0.07\\
0.20 & 0.35 & 0.12 & 0.07\\
0.15 & 0.30 & 0.16 & 0.03\\
\hline
\end{tabular}
\caption{Values of dark fluid effective parameters favored by the CMB angular spectrum peak positions, for $h =0.7$, $n=1$, $\Omega_b^0 = 0.049$ and $\Omega_r^0=9.89 \times 10^{-5}$.\label{table}}
\end{center}
\end{table*}%
However, the allowed values of the different parameters are correlated, as can be seen in Table~\ref{table}. Indeed, for large values of $\tilde{\Omega}_D^0 > \Omega^{ls}_{Dm}$, the permitted values of $\overline{\omega}_D$ are negative, which is compatible with the idea that our fluid behaves today like a cosmological constant whereas it could have behaved mainly like matter at last scattering. For $\tilde{\Omega}_D^0 < \Omega^{ls}_{Dm}$, $\overline{\omega}_D$ is positive, so that the density of dark fluid should decrease more rapidly than a matter density after last scattering. Also, the value of $\overline{\Omega}^{ls}_{Du}$, which can be as much as 0.10, shows that before recombination, the fluid could behave differently from matter, similar to radiation for example.\\
\\
In the standard cosmological model, the two dominant densities after recombination, namely dark matter and dark energy densities, are known to have $\omega_{dm}=0$ and $\omega_{de}<0$, so that it seems safe to impose for the dark fluid $\overline{\omega}_D<0$. In this case, since the negative pressure imposes a deceleration of the dark fluid energy density decrease, we have necessarily $\tilde{\Omega}_D^0 < \Omega_D^0$. In this context, combining the constraints of Eqs. (\ref{resultSN}) and (\ref{resultCMB}) leads to
\begin{eqnarray}
&\Omega_D^0 = 1.005 \pm 0.006  & \omega_D^0 = -0.80 \pm 0.12 \;\;\;\;\; \omega_D^a =  0.9 \pm 0.5 \nonumber\\
&\Omega^{ls}_{Dm} = 0.35 \pm 0.05 & \overline{\Omega}^{ls}_{Du} = 0.065 \pm 0.025\label{resultcombined}\\
& \overline{\omega}_D = -0.08 \pm 0.08 & \tilde{\Omega}_D^0 = 0.70 \pm 0.30\nonumber
\end{eqnarray}
Hence, even without the specification of a precise model, the combination of the cosmological observations already provides important constraints on the dark fluid models. With the specification of a precise model, all the features of CMB spectrum should be reproduced, which will lead to more stringent constraints.
\section{Big-Bang Nucleosynthesis}
\noindent Analyses of the big--bang nucleosynthesis (BBN) \cite{BBN1} indicate a discrepancy between the value of the baryonic density calculated from the observed Li and $^4$He abundances, and the one calculated with the observations of deuterium. If the Universe is dominated by radiation at BBN time, the main constraint is that the dark fluid density should be small in comparison to the radiation density and not perturb much the expansion rate after BBN, that is, if one assumes that the dark fluid behavior does not change violently during BBN, the equation of state of the dark fluid around the time of BBN has to be $\omega_D(\mbox{BBN}) \geq 1/3$ with a density smaller than the radiation density, or that its density was completely negligible before BBN. In the case of a real radiative behavior $\omega_D(\mbox{BBN}) = 1/3$, the dark fluid behaves like extra--families of neutrinos, and its density can be constrained \cite{BBN2}:
\begin{equation}
\rho_D(\mbox{\small BBN}) < \frac{7}{8} \left( \frac{4}{11}\right)^{4/3} \frac{\pi^2}{15} \ T^4_{\mbox{\small BBN}} \approx 3 \times 10^{-2} (\mbox{MeV})^4 \;\;.
\end{equation}
\section{Local scales}
An important question remains, how to interpret the dark matter problem at local scales and could the dark fluid account for the excess of gravity inside local structures? I provide here a qualitative analysis of how a fluid with a negative pressure at cosmological scale can have an attractive effect at local scale, such as it is observed in galaxies -- for example, with the rotation curves of spiral galaxies \cite{rotationcurves}.\\
\\
In the quasi--Newtonian limit of general relativity, the gravitational potential reads:
\begin{equation}
\Phi\left( \vec{r} \right) \; = \; - \, 2 \, G \; {\displaystyle \int} \, {\displaystyle \frac{S_{00}\left(\vec{r}\,'\right)}{|\,\vec{r}\,'\,-\,\vec{r}\,|}}\,d^3\vec{r}\,' \;\; ,
\end{equation}
where $S_{\mu\nu}$ is the source tensor,  {\it i.e.} the energy-momentum tensor minus its trace. For a fluid at rest, with an equation of state $P=\omega \rho$:
\begin{equation}
S_{00}=\frac{1}{2} \rho (1+3 \omega) \;\;,
\end{equation}
and consequently the fluid has an attractive effect if $\omega > -1/3$. From the study of supernov\ae~it seems that our dark fluid is not in that state today, so that its effects are mainly repulsive. Nevertheless, it is possible that dark fluid has a different behavior on cosmological and on local scales. Indeed, the density and pressure of the fluid on cosmological scale are spacial averages of the local density and pressure, and one can assume that:
\begin{eqnarray}
\nonumber \rho\left(t,\vec{r}\right) &=& \rho^{\mbox{cosmo}}\left(t\right) + \delta\rho\left(t,\vec{r}\right)\;\;,\\
P\left(t,\vec{r}\right) &=& P^{\mbox{cosmo}}\left(t\right) + \delta P\left(t,\vec{r}\right) \;\;,
\end{eqnarray}
where $\rho^{\mbox{cosmo}}$ and $P^{\mbox{cosmo}}$ are the cosmological density and pressure, with the spacial averages:
\begin{equation}
<\delta\rho\left(t,\vec{r}\right)>=<\delta P\left(t,\vec{r}\right)>= 0 \;\;.
\end{equation}
The density of dark fluid at cosmological scales today is of the order of the critical density,  {\it i.e.} $\rho^0_c\approx~9\times~10^{-29}\mbox{g.cm}^{-3}$. One can compare it to the estimated matter density in the Milky Way at the radius of the Sun $\rho^{\mbox{Sun}} \approx 5 \times 10^{-24}\mbox{g.cm}^{-3}$ \cite{milky_way}. Hence, even if the dark fluid local density would represent 1\% of this total matter local density, its value would have been much higher than the cosmological densities today. Therefore, at local scales, one can assume that $\delta\rho\left(t,\vec{r}\right) \gg \rho^{\mbox{cosmo}}\left(t\right)$ and consequently write:
\begin{equation}
S_{00} \approx \frac{1}{2} (\delta\rho + 3 \delta P) \;\;.
\end{equation}
To have a net attraction, we get finally a constraint  similar to the previous ones, $\delta\rho > -3 \delta P$, but in this case we do not have to use the cosmological constraints because the local behavior of the dark fluid can be very different from the cosmological one. In the standard cosmological scenario, the only components which have a gravitational influence on local scales are baryonic matter and dark matter. The dark fluid should account at local scales for the dark matter behavior. The most successful dark matter models involve \textit{cold} dark matter, which can be considered as a pressureless fluid. Thus, we can consider that a locally pressureless dark fluid is more likely to success in describing local scales, and it is safe to assume that $\delta P$ is negligible at local scales. As a consequence, the local behavior of the dark fluid is matter-like, and that the usual Newtonian equation can be retrieved:
\begin{equation}
\Phi\left( \vec{r} \right) \; = \; - \, G \; {\displaystyle \int} \, {\displaystyle \frac{\delta\rho\left(\vec{r}\,'\right)}{|\,\vec{r}\,'\,-\,\vec{r}\,|}}\,d^3\vec{r}\,' \;\; .
\end{equation}
This local behavior would of course have to be verified quantitatively for each dark fluid model, but give nevertheless the possibility to have a unified explanation at all scales.\\
\section{Scalar Field Dark Fluids}
\begin{figure}[h]
   \centering
   \includegraphics[width=6.cm,angle=270]{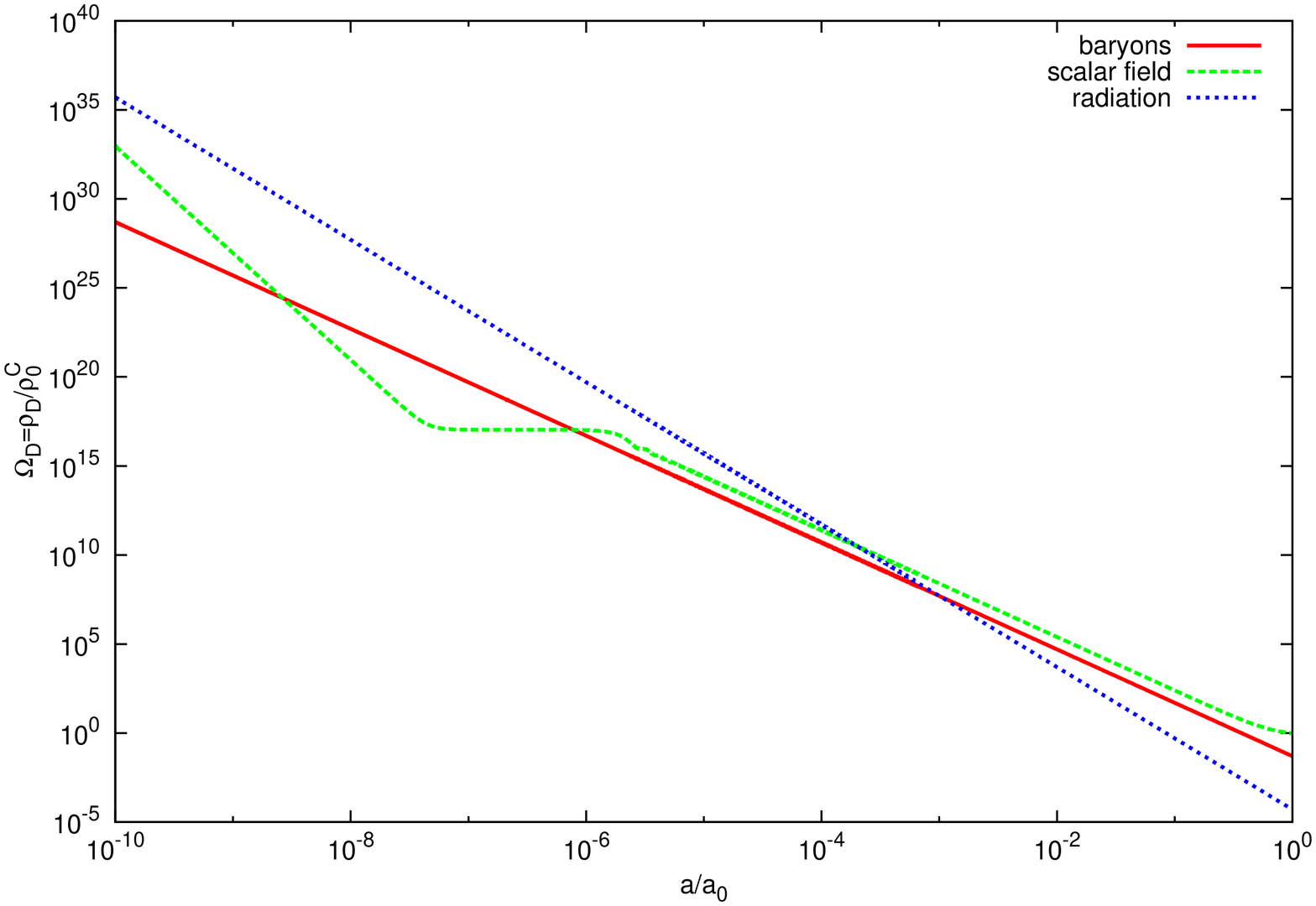}\includegraphics[width=6.cm,angle=270]{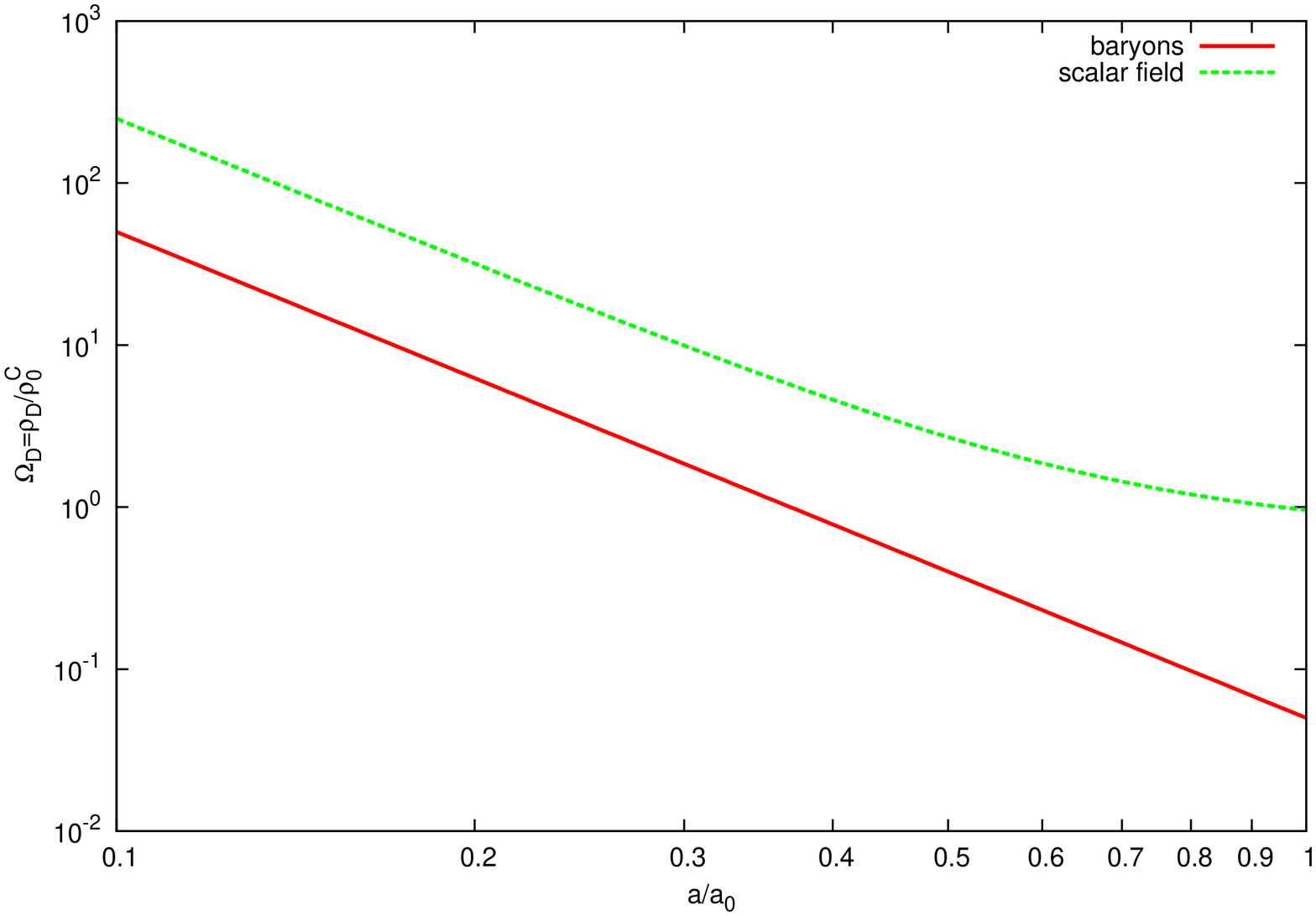}
   \caption{Cosmological evolution of the density of a complex scalar field (dark fluid) in comparison to the densities of baryonic matter and radiation.}
   \label{cosmo}
   \end{figure}

\noindent In the literature, only few fluids behaving similar to a dark fluid are considered. One can also note that the generalized Chaplygin gas \cite{chaplygin}, based on D-brane theories, seems in agreement with the cosmological constraints established in this article. A further analysis is still needed, in particular concerning the growth of structures with a dominant Chaplygin gas density, or the local behavior of such a fluid.\\ 
Here we consider a different model based on a scalar field \cite{padmanaban,arbey3}. Indeed, scalar fields are very useful in explaining the behavior of the dark energy today \cite{quintessence1,quintessence2}, and recent analyses have shown that they can behave like matter at local scales \cite{arbey1,sflocal} as well as at cosmological scales \cite{arbey2,sfcosmo}.\\
Let us therefore consider a real scalar field associated with a Lagrangian density
\begin{equation}
\mathcal{L} \; = \; g^{\mu \nu} \, \partial_{\mu} \varphi \, \partial_{\nu} \varphi \; - \; V \left( \varphi \right) \;\; .
\end{equation}
The pressure and the density of such a field at cosmological scales are given by:
\begin{eqnarray}
P_\varphi =\frac{1}{2}\dot\varphi^2 - V\left( \varphi \right) \;\;,\;\; \rho_\varphi =\frac{1}{2}\dot\varphi^2 + V\left( \varphi \right) \;\;.
\end{eqnarray}
Therefore, the pressure is negative if the potential dominates, and negligible if the potential equilibrates the kinetic term. Thus, a scalar field can be a good candidate for the dark fluid if it respects in particular the following constraints:\\
\indent -- its density at the time of the BBN decreases at least as fast as the density of radiation,\\
\indent -- its density from the time of last scattering to the time of structure formation evolves nearly like matter, and so $\frac{1}{2}\dot\varphi^2\approx~V\left(\varphi\right)$,\\
\indent -- after the growth of perturbations, because the scalar field dominates the Universe, its potential does not equilibrate the kinetic term anymore and  begins to dominate, leading to a cosmological constant behavior in the future.\\
\\
The main parameter of such a model is the same as that of quintessence models: the potential. In quintessence models with real scalar fields, one looks for potentials which provide a cosmological constant--like behavior today, and decreasing potentials seem to be favored. If one considers now complex scalar fields, it was shown in \cite{arbey2} that such fields can behave like cosmological matter if their potentials contain a dominant term in $m^2 |\phi|^2$. Therefore, a way to find a ``good'' potential would be to consider a superposition of a decreasing potential which would begin to dominate today, and of the increasing quadratic term which has to dominate at least until structure formation and can nevertheless lead to an attractive effect at local scales today. In that case, the simplest potential matching to the necessary properties is:
\begin{equation}
V(\varphi) =m^2 \varphi^\dagger \varphi + A \exp(-B |\varphi|) \;\;.
\end{equation}
Imposing $m \approx 10^{-23}$ eV \cite{arbey2} and choosing $A \approx \rho_{de}^0$ and $B$ such as $A \exp(-B |\varphi_0|) \sim m^2 \varphi_0^\dagger \varphi_0$ today, this potential leads to the cosmological behavior of figure \ref{cosmo}.
As can be infered from the figure, during structure formation the quadratic part of the potential dominates and the field behaves like matter. Later on, a phase transition has occured when the second part of the potential began to dominate, and the dark fluid has undertaken a quintessence behavior. The case described in the figure is consistent with the constraints presented in the previous sections, and the Hubble constant and the age of the Universe obtained in this model are very close to the ones from the standard model, and therefore are compatible with observations. It is also the case for the potential described in \cite{arbey3}. \\
\\
As developed in \cite{arbey3}, such a scalar field can also explain the excess of gravity at local scales. To do so, let us imagine that the Universe is filled with the scalar field. Where -- and when -- the density of baryonic matter is large, the scalar field would, through gravitational interaction, get a large kinetic term which can equilibrate its potential, giving the field an attractive net force at local scales; where baryons have a very low density, the field can be at rest with a dominating potential, providing repulsion. Hence, at local scales, where the baryon density is high, the field behaves like matter. Where the baryon density is small, {\it i.e.} away from galaxies and clusters, the gravitational interaction is not strong enough to increase the kinetic term of the scalar field, so that the potential dominates, and one can then observe the effects of a negative pressure. In the past, baryons were uniformly dense, so that the kinetic term was large everywhere, leading to a uniform matter behavior under these conditions. In such a scenario, the local attractive behavior can be in agreement with the cosmological repulsive behavior, and a complete cosmological scenario can therefore be worked out. The precedent studies in this context confirm that such a scenario is plausible \cite{arbey3}, but deeper studies involving structure formation and CMB physics have to be carried out in order to check the validity of the scalar field dark fluid model.
\section{Conclusion and Perspectives}
\noindent Astrophysical and cosmological observations are usually considered in terms of dark matter and dark energy. As we showed here, they can also be interpreted differently, in terms of unifying dark fluids, which could advantageously replace models containing two dark components. In this article, we have derived general constraints on such models in light of recent cosmological data, and we have considered the specific example of scalar field dark fluids and showed that they can be in agreement with the obtained constraints.
\section*{Acknowledgements}
\noindent I would like to thank Farvah Mahmoudi for her comments.

\end{document}